\documentstyle[12pt,epsfig]{article}
\begin{document}
\newcommand{\lsim}{\lower .5ex\hbox{$\buildrel < \over {\sim}$}}
\newcommand{\gsim}{\lower .5ex\hbox{$\buildrel > \over {\sim}$}}

\title{Fragmentation cross sections of 158 A GeV $Pb$ ions in various targets measured with $CR39$ nuclear track detectors.}
\author{S. Cecchini$^1$, G. Giacomelli, M.Giorgini, G. Mandrioli, \\ L. Patrizii, V. Popa$^2$, P. Serra, G. Sirri, and M. Spurio\footnote{corresponding author. E-mail: spurio@bo.infn.it} \\ \\
\small Department of Physics of the University of Bologna and INFN Sezione di Bologna, \\ 
\small Viale Berti Pichat 6/2, I-40127 Bologna, Italy \\
\small $^1$ Also IASF/CNR, Sezione di Bologna, I-40129 Bologna, Italy \\
\small $^2$ Also Institute for Space Sciences, 76900 Bucharest, Romania }

\maketitle
\vspace{-12.0cm} 
\hspace{8.0cm} DFUB 02/01

\hspace{8.0cm} University of Bologna
\vspace{12.0cm} 

\begin{abstract}
We report the measurement of the fragmentation cross sections in high-energy nucleus-nucleus collisions using the 158 A GeV $Pb^{82+}$ beam from the CERN-SPS. The fragments have charges changed from that of the incident projectile nucleus by $\Delta Z=Z_{Pb}-Z_{frag}$, with $8 \le \Delta Z <75$. The targets range from polyethylene to lead. Charge identification is made with $CR39$ nuclear track detectors, measured with an automatic image analyzer system.
The measured fragmentation cross sections are parameterized with an empirical relation in terms of the atomic mass of the target, and of the charge of the final fragment. 

\vspace{0.2cm}
\end{abstract}

\section{Introduction}

Experimental results on the fragmentation properties of high-energy nuclei are relevant for nuclear physics, cosmic ray physics and astrophysics. In particular, the fragmentation cross sections (especially in hydrogen) are needed to understand the cosmic ray propagation in our Galaxy.

In this paper it is studied the production of nuclear fragments with charge changed from that of the incident projectile nucleus by $\Delta Z=Z_{Pb}-Z_{frag}$, with $8 \le \Delta Z<75$.  158 A GeV $Pb$ ions (charge $Z_{Pb}= 82$) from the CERN-SPS were used as projectiles.
Six stacks made of $CR39$ nuclear track detectors and of six target materials ($CH_2$, $CR39$, $C$, $Al$, $Cu$, $Pb$) were used.
The exposures were performed at normal incidence; the total number of lead ions collected in one stack was about $5.8 \times 10^4$. The present analysis refers to a sample of about $3.5 \times 10^4$ incident $Pb$ ions on each target. 
This work completes the analysis presented in a previous paper [1], where we reported on the production of fragments with charge changed by $-1\le \Delta Z < 8$. 

In nucleus-nucleus interactions, central and peripheral collisions via strong interaction play an important role. The electromagnetic interaction can also contribute, as proposed in \cite{bertulani}.
In the present work, the leading outgoing fragment nucleus that keep most of the original longitudinal velocity of the $Pb$ projectile is studied; the fragment is inside a scattering angle smaller than 1 mrad. This exclusive study of the leading fragment is somewhat complementary to the inclusive measurement of all fragments with $Z \ge 7$ studied in the interaction of the same beam projectiles on targets similar to ours \cite{huntrup}.

Our $CR39$ nuclear track detector sheets (chemical composition $(C_{12} H_{18} O_{7})_n$) were manufactured by the Intercast Europe Co. of Parma, Italy, and are similar to those used in a large area experimental search for magnetic monopoles at the Gran Sasso Laboratory (MACRO) \cite{macro}. The response of the detector was discussed in \cite{cecco96}.

\begin{figure}[htb]
\vspace{-0.5cm}
\begin{center}
	\mbox{\hspace{-1cm}
	\epsfig{file=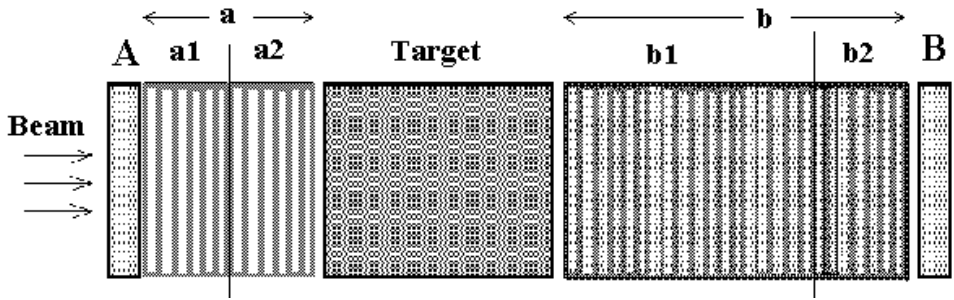,height=5cm}}
\end{center}
\vspace{-0.5cm}
\caption{\small Sketch of a stack used for the fragmentation studies. ``A" and ``B" are 1.4 mm thick $CR39$ sheets. ``a"= ``a1+a2" and ``b" = ``b1+b2" are stacks (about 7 and 15 mm thick, respectively) of layers of 0.6 mm thick $CR39$; the target is typically 10 mm thick.}
  \label{fig:sketch}
\end{figure}

Here we report and discuss the measurements of the fragmentation cross sections $\sigma(\Delta Z,A_T)$ for $\Delta Z$  up to $74$ and for various targets with mass $A_T$. The dependence of the fragmentation cross sections on the target mass and charge of the final fragments is analyzed in terms of an empirical relation. 
We do not detect (because of the $CR39$ threshold) fragments with charge $Z<7$ and we do not consider multi-fragmentation processes.

\section{Experimental procedure}

The leading charged fragments originating from $Pb$ projectile interactions with target nuclei keep most of the projectile longitudinal velocity. They can be identified downstream of the target in $CR39$ detectors.
The detection principle \cite{bb} of the $CR39$ is based on the fact that a throughgoing ionizing particle produces a cylindrical radiation-damaged region along the ion trajectory.
The damaged region of the material becomes chemically reactive and can be etched by an appropriate chemical treatment. As a result, an etched cone is formed on both sides of each detector sheet. The cones are visible under a microscope. For energies larger than $\sim 1$ GeV/nucleon, the base area of the post-etching cones depends on the projectile charge only. 

Each of our six stacks had the composition shown in Fig. 1. Upstream (``A") and downstream (``B") of the target there are 1.4 mm thick $CR39$ foils. 12 sheets of 0.6 mm $CR39$ foils (``a" in Fig. 1) are placed between ``A" and the target; 24 sheets of 0.6 mm $CR39$ foils (``b" in Fig. 1) are added after the target. The $\sim 10$ mm thick targets, consisted of $Pb$, $Cu$, $Al$, $C$, $CH_2$ and $CR39$ sheets. Table 1 gives the target density and thickness, and other technical details. The targets are thin enough to minimize multiple interactions and thick enough to produce a sufficient number of fragments. 

After exposure, the $CR39$ detectors were etched for 268 h in a 6N NaOH water solution at a temperature of $45 ^o$ C. With these etching conditions, the base diameter of a cone produced by an ion with charge $Z \ge 75$ is around $75\ \mu $m, while the smallest base diameter (for $Z=7$) is around $7\ \mu$m  \cite{nim}.

For scanning the detector surfaces and measuring the etch-pit cone areas, an automatic image analyzer system was used \cite{elbek}. The main components of the system are a microscope with a motorized stage, a video camera and a computer. For each etch-pit cone the base area, the eccentricity, the central brightness and the coordinates of the center were measured. 

In this analysis, the six $CR39$ sheets immediately before the target (``a2" region in Fig. 1) and the 14 sheets after the target (``b1" region) were etched and the areas of the pit cones measured (on the front side of each sheet) with the automatic system. A tracking procedure was used to reconstruct the path of the fragments. To identify the projectile and fragment charges we performed an average of the measured etch-pit areas for each track over several sheets. 
An example of the distribution of the etched cone areas averaged over at least 12 $CR39$ sheets located after the fragmentation target (``b1" region) is shown in Fig. 2. With this method, we can separate all single peaks from $Z= 7$ up to $Z=74$. The charge resolution at $Z=10$ is around $0.3e$ for a single measurement and $\sim 0.1e$ for the average of 12 measurements; the resolution worsens slightly with increasing values of $Z$.

The identification of fragments with $Z \ge 75$ requires the measurement of the ``cone heights". The technique was used in \cite{dekhi00}, to measure the total, fragmentation (for $1< \Delta Z < 8$) and pick-up ($\Delta Z =-1$) charge-changing cross sections of 158 A GeV $Pb^{82+}$ ions. 

\begin{table}
\begin{center}
\begin{tabular}
{|c|cccc|} \hline
Target & $A_T$ & $Z_T$ & $\rho_T$ & $t_T$ \\
       &       &       & $g cm^{-3}$ & $cm$ \\ \hline
$Pb$ & 207 & 82 & 11.331 & $0.98 \pm 0.01$ \\
$Cu$ & 63.5 & 29 & 8.901 & $0.99 \pm 0.01$ \\
$Al$ & 27 & 13 & 2.692 & $1.04 \pm 0.01$   \\
$C$ & 12 & 6 & 1.733 & $1.01 \pm 0.01$    \\
CR39 & 7.4 & 4.0 & 1.310 & $1.02 \pm 0.01$ \\
$CH_2$ & 4.7 &2.7 & 0.952 & $1.02 \pm 0.01$ \\ \hline
\end{tabular}
\end {center}
\caption{\small Summary of the properties of the six targets exposed to the relativistic $Pb$ beam. $\rho_T\ (g cm^  {-3})$ is the density, $A_T$ and $Z_T$ the atomic mass and charge of the target materials, and $t_T\ (cm)$ is the target thickness.}
\label{tab:target}
\end{table}

\section{Determination of the total and partial charge changing cross section}

The interaction of the beam ions with the target nuclei yields nuclear fragments with a final charge $Z_{frag}$. After the target, only the leading fragment, collinear with the incident $Pb$ nucleus (angular difference less than $1$ mrad), is measured. 

In \cite{dekhi00}, we measured the total charge-changing cross sections, $\sigma_{tot}$; the data were fitted with the cross section model discussed in \cite{pri,sam}: 
\begin{equation}
\sigma_{tot}= \sigma_{nucl} + \sigma_{EMD} = a(A_P^{1/3}+A_T^{1/3}-b)^2+\alpha Z_T^{\delta}~ \label{eq:fit}
\end{equation}
The first term of eq. 1 represents the nuclear contribution to the cross section, while the second is the electromagnetic term. (The values of the parameters $a$ and $\delta$, and the fitted values of $b$ and $\alpha$ were given in \cite{dekhi00}).

The nuclear term is nearly energy-independent; it contains an explicit dependence on the atomic mass number $A_T$ of the target and $A_P$ of the projectile. 
The electromagnetic term (electromagnetic dissociation, EMD) is important for the emission of single nucleons or $\alpha$ particles from a projectile, when interacting with the electromagnetic field of a heavy nucleus. It has been shown in \cite{hir} that this term depends on the beam energy.

\begin{figure}
\vspace{-5.2cm}
\begin{center}
	\mbox{\hspace{9cm}
	\epsfig{file=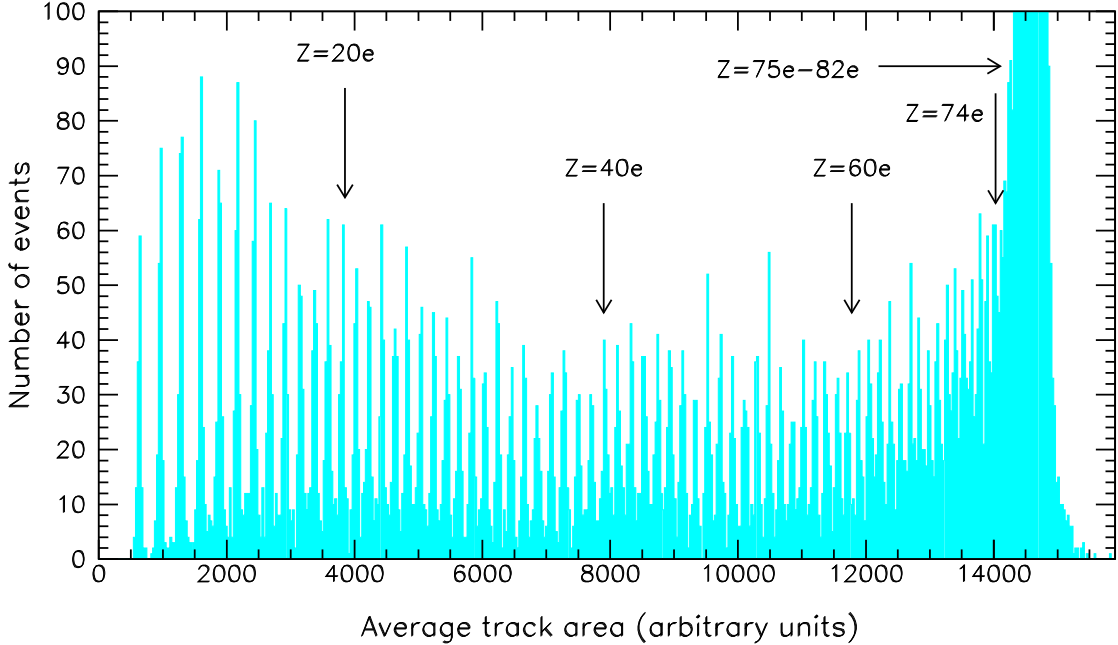,height=3.0cm,angle=270}}
\end{center}
\vspace{-0.5cm}
\caption{\small Distribution of etched cone base areas averaged over at least 12 faces of the $CR39$ sheets located after the Al target (region ``b1" of Fig. 1) and measured with our automatic image analyzer system. This procedure yields a good charge resolution for fragments with $Z \leq 60$ and an acceptable one up to $Z \sim 74$. With this method, the fragment tracks with $Z \geq 75$ and the lead tracks cannot be distinguished, and the ``cone height'' method (described in [1]) was used.}
\label{fig:coni}
\end{figure}

Our measured total charge-changing cross sections $\sigma_{tot}$ obtained in the collision of 158 A GeV $Pb$ ions are given in the second column of Table 2. In the same table, the values of the nuclear (column 3) and electromagnetic (column 4) contributions, estimated from eq. 1, are included. 

The contributions from the $\sigma_{nucl}$ term are in good agreement with what reported in \cite{hir,ger} using 10.6 GeV/nucleon $Au$ ions nuclei. It is to be mentioned that our values of $\sigma_{tot}$ from targets with the heaviest masses (Cu,Pb), are about 30-50\% higher than measured at lower energies \cite{pri,sam,hir,ger}. This difference is ascribed to the electromagnetic term $\sigma_{EMD}$: in our case, this contribution is dominant ($\sim 6800$ mb) for $Pb-Pb$ interactions, and it is $\sim 20\%$ of $\sigma_{tot}$ for $Pb-Cu$. In the case of 10.6 GeV/nucleon $Au$ ions on $Pb$ target, the estimated $\sigma_{EMD}$ is around $350\ mb$. From the same calculations (see Fig. 3 of \cite{hir} and Fig. 13 of \cite{ger}), the electromagnetic dissociation is expected to play a substantial role at our energies (158 GeV/nucleon) for $Pb-Pb$ collisions.

Hydrogen cross sections were obtained from the measured cross sections on Carbon and on $CH_2$ using the formula:
\begin{equation}
\sigma_H = {1 \over 2} (3 \sigma_{CH_2} -  \sigma_{C}) 
\end{equation}
The total cross section on the H target is slightly higher (at the level of $\sim 2\sigma$) than reported in \cite{hir,ger}. In this case, the EMD contribution is negligible. 

The partial cross sections for fragments with final charge $75 \le Z \le 81$ ($\sigma_{75-81}$) measured in \cite{dekhi00} are also given in Table 2 (column 5). 
The fragments with charge $Z<7$ were not measured because of detection threshold of our $CR39$. The contribution to the total cross section from  the fragmentation of lead ions into fragments with $Z<7$ was estimated \cite{dekhi00} as $\sim$ 10\% of the partial cross section for fragments with $7 \le Z \le 74$ ($\sigma_{7-74}$).
The $\sigma_{7-74}$ (column 6 of Table 2) were derived as the difference between $\sigma_{tot}$ and $\sigma_{75-81}$, taking into account the contribution on $\sigma_{tot}$ from fragments with charge $Z<7$.

The measured values of $\sigma^{new}_{7-74}$ obtained in this work are given in the last column of Table 2; they are obtained by summing the contribution of each charge channel as described in the next section. 
The old and the new values agree within errors: this represents a consistency check of the two experimental techniques. 

The errors quoted in Table 2 are only statistical, and they are largely reduced in the new measurements with respect to the old ones. For the old $\sigma_{7-74}$, it was estimated an additional systematic error of 10\%. For the new $\sigma^{new}_{7-74}$, we estimate a systematic error of 8\%, as discussed in the next section.

\begin{table}
\begin{center}
\begin{tabular}
{c|ccc|c|cc} \hline
Target & $\sigma_{tot}\ (mb)$ & $\sigma_{nucl}$ & $\sigma_{EMD}$ & $\sigma_{75-81}\ (mb)$ & $\sigma_{7-74}\ (mb)$ & $\sigma^{new} _{7-74}\ (mb)$ \\ 
& [1]  & (mb) & (mb)&  [1] &  [1] &  $This\ work$ \\\hline 
$Pb$ &$12850\pm 640$ & 6357 & 6794 & $7460\pm 360$ & $4850 \pm 730$ & $4460 \pm 140$ \\
$Cu$ &$5090\pm 280$ & 4315 & 943 & $1280\pm 110$ & $3430 \pm 300$ & $3720 \pm 110$ \\
$Al$ & $3800\pm 170$ & 3406 & 205 & $1080 \pm 120$& $2450 \pm 200$ & $2300 \pm 70$ \\
$C$  & $2910\pm 210$ & 2825 & 47 & $510 \pm 70$ & $2160 \pm 220$ & $2250 \pm 70$ \\
CR39 & $2640\pm 80$ & 2565 & 22 & $450 \pm 20$ & $1980 \pm 100$ & $ 2060\pm 60$ \\
$CH_2$&  $2270\pm 150$ & 2366 & 10 & $800 \pm 70$ & $1320 \pm 170$ & $1290 \pm 40$ \\
\hline
$H$& $1950\pm 280$ & 1909 & 2 &$900 \pm 180$ & $940 \pm 330$ & $810 \pm 180$ \\
\hline
\end{tabular}
\end {center}
\caption{\small Total and partial charge-changing cross sections for 158 GeV/nucleon $Pb$ ions on various targets. The $Pb$ total charge-changing cross sections $\sigma_{tot}$ and the partial cross sections for fragmentation into nuclear fragments with charge $Z$ from 75 up to 81 ($\sigma_{75-81}$) were measured with a different charge identification technique [1]. The nuclear and the electromagnetic contributions (columns 3 and 4) to the total cross sections were obtained using eq. 1. The last column gives the improved measurements of $\sigma_{7-74}$ with the method discussed in this paper. The quoted errors include only the statistical uncertainties.}
\label{tab:simu}
\end{table}

\section{Fragmentation charge changing cross sections}
In this work, we require the presence of a reconstructed track both before and after the target. A reconstructed track must have at least 3 measured $CR39$ faces in the six sheets immediately before the target (in the $CR39$ region ``a2" of Fig. 1), and at least 3 measured faces immediately after the target (region ``b1"). Each measured etch-pit area must be spatially correlated with the others to form a track. This procedure leads to high efficiencies on the number of reconstructed fragments, while preserving good charge identification.

Each track must be classified as a ``Pb" one before the target.
With the present method of charge identification, all nuclei with $Z \ge 75$ are considered as ``Pb" in the analysis. The contamination does not affect in a significant way our measurement of the fragmentation cross-sections from the pure lead ions, since there is a small difference on the $\sigma_{tot}$ for nuclei with similar atomic mass $A_P$ (see eq. 1). 
The $Pb$ beam has an additional 1\% contamination from ions with charge $Z=81$.
Fragments with $7 \le Z \le 74$ produced in the first 3.5 mm of $CR39$ (''a1" region of Fig. 1) were identified in the ''a2" region and excluded for further analysis. 

The fragmentation charge-changing cross section for the $A_P=Pb$ projectile nuclei was evaluated using the formula \cite{bha}
\begin{equation}
\sigma(\Delta Z,A_T) = \sigma^T_{i} = {1 \over x_T} {N^T_i \over N^T_0 } \end{equation}
where $N^T_i$ is the number of nuclear fragments with charge $Z_i$, with $i=7-74$, produced in the target; $N^T_0$ is the number of non interacting beam ions (including all nuclei with $75 \le Z \le 81$). Both $N^T_i$ and $N^T_0$ are identified after the target, in the region ``b1" of Fig. 1.  
For each target T, $x_T=\rho_T \cdot t_T \cdot N_A/A_T (mb^{-1})$ depends on  the target density $\rho_T$, atomic mass $A_T$ and thickness $t_T$ (see Table 1); $N_A$ is the Avogadro number.
In this procedure, successive fragmentation processes are neglected. However, using the measured $\sigma_{tot}$ of Table 2, we estimate to have at most 5\% of multiple fragmentation in the Cu target, and 2\% in the remaining targets.  

Both measured numbers $N^T_i$ and $N^T_0$ in eq. 3 must be corrected to take into account the probability that some of the $N^T_i$ fragments were produced in the 3.5 mm $CR39$ layer (``a2") before the target. The correction can be estimated using the measurements of the partial cross sections for the production of fragments ($\sigma^{CR39}_{7-74}$) in the pure $CR39$ target. 
As an example of the correction, in the case of the $C$ target, we estimate that 7.3\% of the incoming $Pb$ ions interact upstream of the target in the $CR39$ ``a2" region. This contributes for the 29\% of the total number of fragments with $7 \le Z \le 74$ measured downstream of the target. 
We estimate at most a 4\% systematic uncertainty from this correction procedure, because we assume equal probability for the production of fragments with charge $7 \le Z \le 74$.

We show in Fig. 3 the values of the fragmentation cross sections for each target and for each fragment. The data have an average statistical error of $\sim 10$\%. The estimated total systematic uncertainty, arising mainly from the neglect of multiple fragmentations, from the identification of the charge $Z$ of the fragment, from the correction procedure and from the contamination of the beam, is about 8\%. 

The hydrogen fragmentation cross-sections (shown in Fig. 5) were estimated using eq. 2; in this case, the overall uncertainty is about 25\%.

\begin{figure}
\vspace{-0.5cm}
\begin{center}
	\mbox{\hspace{1.5cm}
	\epsfig{file=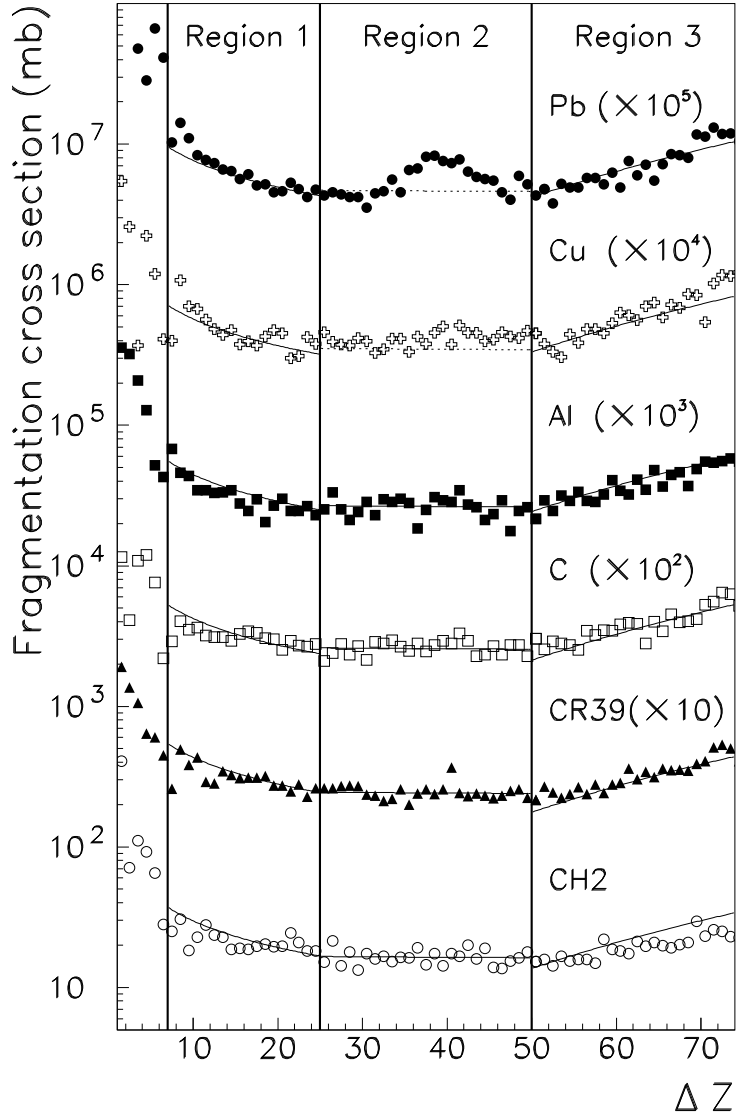,height=16cm}}
\end{center}
\vspace{-0.5cm}
\caption{\small Fragmentation charge-changing cross sections (in mb) for 158 A GeV $Pb$ nuclei incident on six targets, as a function of the charge change $\Delta Z$. For display purposes, for a given target the cross-sections are multiplied by the number indicated next to the target symbol. The lines are the fits of the cross section data to the power laws (eq. 5) with the parameters given in Table 3. Data from the $Pb$ and $Cu$ targets in Region 2 were not used in the fit.}
  \label{fig:cross}
\end{figure}

\vspace{-0.2cm}
\section{Parameterization of the fragmentation cross sections.} 
The partial cross sections for nuclear fragmentation $\sigma(\Delta Z,A_T,A_P)$ of an $A_P$ projectile on a target $A_T$, have been shown to be factorizable \cite{bre9}
\begin{equation}
\sigma{(\Delta Z,A_T,A_P})= \gamma_{PT} \cdot \gamma_{\Delta Z} 
\label{eq:frag}
\end{equation}
where $\gamma_{\Delta Z}$ depends on the projectile and on the fragment, and $\gamma_{PT}$ is the target factor which is expected to be proportional to $A_T^{1/3}$. The $\gamma_{\Delta Z}$ term contains both a nuclear and an electromagnetic contribution. The electromagnetic one is expected to be sizeable only for ${\Delta Z}$ of the order of unity, and is neglected in the following discussion.

As shown in Fig. 3, for a fixed value of ${\Delta Z}$, the fragmentation cross-sections increase with the atomic target mass (contribution of the $\gamma_{PT}$ term in eq. 4). This effect is true for all values of ${\Delta Z}$. For a fixed target, a single power law cannot represent all fragmentation cross-section data.
Three different regions have to be considered: 
\begin{enumerate}
\item $7 \lsim {\Delta Z} \lsim 25$: the cross sections decrease with increasing ${\Delta Z}$;
\item $25 \lsim {\Delta Z} \lsim 50$: the cross sections are almost constant (with the exception of the $Pb$ and $Cu$ targets);
\item $50 \lsim {\Delta Z} \lsim 74$: the cross sections increase with increasing ${\Delta Z}$;
\end{enumerate}

For each of the three regions, we fit the measured fragmentation cross sections with a power-law relation
\begin{equation}
\sigma{(\Delta Z,A_T})= K_j A_T^{\mu_j}  \cdot \Delta Z^{\nu_j}
\label{eq:frapar}
\end{equation}
where $K_j$, ${\mu_j}$ and $\nu_j$ ($j=1,3$) are three free parameters, different for each of the three regions defined above. The term $ A_T^{\mu_j}$ parameterize the target factor $\gamma_{PT}$ of eq. 4, and $\Delta Z^{\nu_j}$ the charge term $\gamma_{\Delta Z}$.

The simple formula (eq. 5) fits well our fragmentation cross-section data in the three $\Delta Z$ intervals and for all targets (apart for the $Cu$ and $Pb$ targets in region 2, as discussed below. These fragmentation cross-section data were not used in the fits to eq. 5). The best-fit parameters are given in Table 3, with the $\chi^2$ and the number of degrees of freedom (DoF). For the $\chi^2$ calculation, each value of the fragmentation cross-section has its own statistical error added in quadrature to the 8\% estimated systematic uncertainty. The uncertainties on the fitted parameters are obtained by requiring that the reduced $\chi^2$ is increased by a unit value. 

For the highest target masses (Pb and Cu) in region 2, the measured cross sections are higher than what expected from eq. 5. For fragments with masses $\sim 1/2$ that of the projectile beam, there is a visible peak structure, especially for the lead target. This feature is probably caused by electromagnetically induced nuclear fission; it was also observed in \cite{huntrup}. The enhancement was also predicted from the processes of $\Delta$-isobar excitation and multiple pion production in \cite{pshe}.

Fig. 4 shows a histogram of the ratios of the measured to calculated (with eq. 5) fragmentation cross-sections. 
The values corresponding to the hydrogen target (which were deduced from the $C$and $CH_2$ targets, and suffer from larger uncertainties) are separately shown in the shadowed histogram. Also in this case, for the calculated fragmentation cross-sections, we used the fit parameters of Table 3.

\begin{table}
\begin{center}
\begin{tabular}
{|lc|ccc|c|} \hline
$j$ & $\Delta Z$ & $K_j$ & $\mu_j$ & $\nu_j$ & $\chi^2$/DoF \\
 & range & $(mb)$ & & & \\ \hline
1 & $7\le \Delta Z \le 25$ & $86 \pm 10$& $0.25 \pm 0.04$ & $-0.63\pm 0.05$ & 154/116 \\
2 &$25 <{\Delta Z}<50$& $12.5\pm 1.7$&$0.29\pm 0.05$ & $-0.03\pm 0.04$ & 140/107   \\
3 & $50 \le \Delta Z \le 74 $  & $(9\pm 1)\times 10^{-4}$ & $0.33 \pm 0.03$ & $2.33\pm 0.03$ & 207/128 \\ \hline
\end{tabular}
\end {center}
\caption{\small Results of the fits to eq. 5 in three $\Delta Z$ intervals. The measured fragmentation cross sections on all targets ranging from $CH_2$ up to lead were used, except for the $Cu$ and the $Pb$ in region 2.}
\label{tab:target}
\end{table}

In the three regions, the exponent $\mu$ of the atomic mass target $A_T$ in eq. 5 is compatible with the value $1/3$ expected from a simple geometrical model. In a recent paper \cite{wadd} a weaker dependence was found, but with a different parameterization for the cross sections. Their data were also fitted as function of the energy $E$ of the incoming beam, as $\sigma \sim E^{n(A_T,\Delta Z)}$.

The measurement of the dependence of the fragmentation cross-sections versus $\Delta Z$ values up to 74 is the main feature of this work.
In region 1 ($\Delta Z \lsim 25$), we can compare our results with those obtained by \cite{ger}, using 10.6 A GeV $Au$ projectiles and different targets. Here, the $\nu_1$ coefficient of the $\Delta Z$ variable ranges between -0.58 and -0.62. Because of the agreement with our measured value ($-0.63\pm 0.05$) at 158 GeV A, we can conclude that at high energy $\nu_1$ is almost energy-independent.
In region 2, there is almost no dependence on $\Delta Z$: our fitted coefficient value $\nu_2$ is consistent with zero, with the exception of the discussed case of the highest target masses.
Finally, there is an increases of $\sigma(\Delta Z, A_T)$ with $\Delta Z$ for the production of fragments with small masses (region 3, $\Delta Z \gsim 50$).

\begin{figure}[htb]
\vspace{-2.5cm}
\begin{center}
	\mbox{
	\epsfig{file=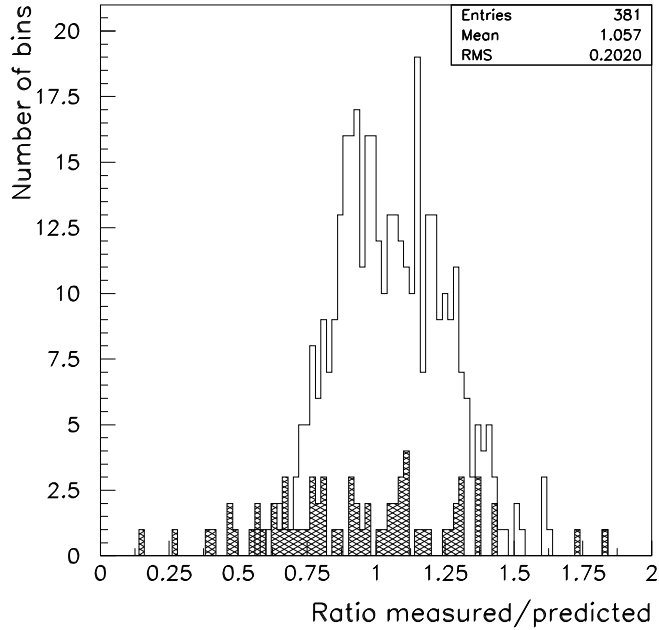,height=10cm}}
\end{center}
\vspace{-0.5cm}
\caption{\small Histogram of the ratios of measured to predicted fragmentation cross-sections, for all targets from $CH_2$ to $Pb$. The predicted values of $\sigma{(\Delta Z,A_T})$ were evaluated with eq. 5 and the parameters given in Table 3. The shadowed histogram corresponds to the hydrogen case.}
  \label{fig:coni}
\end{figure}

\section{Hydrogen fragmentation charge changing cross section}
From data on the $CH_2$ and $C$ targets, we computed with eq. 2 the fragmentation and the total charge-changing cross-sections on Hydrogen.
The case of the Hydrogen target is of particular interest for the astrophysical problem of cosmic rays propagation in our Galaxy. In fact, as discussed long ago \cite{sie}, the composition of nuclei in the cosmic radiation is changed by the interactions with the interstellar hydrogen medium. Recent data on the fragmentation cross sections using relativistic gold nuclei have been presented up to $\Delta Z=30$ \cite{waddh}.

Our fragmentation charge-changing cross sections of 158 GeV/nucleon $Pb$ nuclei on a hydrogen target are shown in Fig. 5. The $\sigma^H(\Delta Z)$ $vs \ \Delta Z$ has the same behavior of that on higher mass targets. For $\Delta Z \lsim 30$ the hydrogen cross sections decrease with increasing $\Delta Z$, and then become almost $\Delta Z$ independent (the increase of $\sigma^H(\Delta Z)$ at large $\Delta Z$ is not observed). The power-law (eq. 5), fits well also the Hydrogen data in regions 1 and 2 ($\chi^2/DoF = 69/45$); the fit is shown in Fig. 5 as a full line.  We used the parameters given in Table 3.

\begin{figure}[htb]
\vspace{-4.5cm}
\begin{center}
	\mbox{
	\epsfig{file=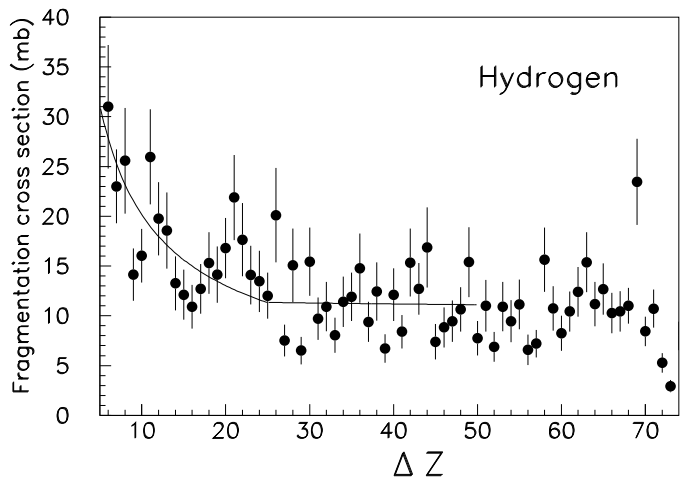,height=10cm}}
\end{center}
\vspace{-0.5cm}
\caption{\small Fragmentation charge-changing cross sections (in mb) for 158 A GeV $Pb$ nuclei on the hydrogen target, as a function of the charge change $\Delta Z$. The line is the prediction of the power law (eq. 5) with the fitted parameters obtained with higher mass targets (see Table 3).}
  \label{fig:sigmah}
\end{figure}

Our partial cross-sections for the $Pb-H$ case, for the production of fragments with charge changed by $1 < \Delta Z < 30$, are about 15\% smaller compared with those reported in the recent measurements of \cite{ger,wadd,waddh}. In this case, $Au$ ions beam of 10.6 GeV/nucleon were used. The decreases of the fragmentation cross sections with increasing energy in the quoted $\Delta Z$ range on the hydrogen target was expected: see for example the Fig. 12 of \cite{wadd}.
Our fragmentation cross section data are also $\sim$ 35\% smaller than those calculated using the semi empirical codes of \cite{sie}, which were modeled from lower energy data.

\section{Conclusions}

We measured the fragmentation charge changing cross sections $\sigma(\Delta Z,A_T)$ using a beam of $Pb$ nuclei of 158 GeV/nucleon and $CR39$ nuclear track detectors. We studied the leading fragment produced in nucleus-nucleus interactions, with charge changes of $8 \le \Delta Z \le 74$. These measurements complement the measurement of \cite{dekhi00}, in which the partial cross sections for $ \Delta Z < 8$ and the pick-up cross sections were measured. This is the first measurement of fragmentation over such a wide interval of $\Delta Z$.

Our data are in agreement with those reported in similar experiments at lower energies, using a 10.6 GeV/nucleon $Au$ beam, when we take into account the increase of the cross sections due to electromagnetic processes. 

The fragmentation cross sections $\sigma(\Delta Z,A_T)$ for the six targets used in this experiment ($Pb,Cu,Al,C,CH_2,CR39$) are presented in Fig. 3. It is possible to have access to the full data table from the URL: 

\noindent $http://www.bo.infn.it/\sim spurio/cr39/table.html$. 

The measured cross sections were parameterized with a simple power-law relation in terms of the mass of the atomic target, and of the charge of the final fragments. A three parameters power law, fits relatively well all our data in each of the three regions of $\Delta Z$. We find an enhancement of the cross sections with respect to this parameterization, for the production of fragments with $\Delta Z \sim 40$ in the collisions of the $Pb$ ions with the highest mass targets (especially in the $Pb$-Pb case). This enhancement is probably due to electromagnetically induced nuclear fissions. 

For all values of the produced fragment charge, the fragmentation cross section depends on the target atomic mass as $\sim A_T^{1/3}$. This means that also for these very high kinetic energies, the interaction depends on the target nuclear radius, at least for the leading fragments.

From data on $CH_2$ and $C$ targets, we derived the fragmentation cross sections on Hydrogen. These cross sections up to $\Delta Z \lsim 50$ follow the same power-law parameterization used for higher mass targets. Our fragmentation cross-sections for high-energy projectiles are lower that those measured with the 10.6 GeV/nucleon $Au$ beam, and significantly lower than those estimated by the semi-empirical calculations used for cosmic rays propagation.

\vspace{20 mm} 
{\bf Acknowledgements}
\vspace{2 mm} 
 
We thank the CERN SPS staff for the good performance of the $Pb$ beam and for the exposures. We acknowledge the contribution of Ashavani Kumar and of our technical staff: E. Bottazzi, L. Degli Esposti, D. Di Ferdinando, R. Giacomelli, C. Valieri and V. Togo. V. P. thanks INFN for hospitality and financial support.

\end{document}